\documentclass[twocolumn,showpacs,preprintnumbers,amsmath,amssymb]{revtex4}

\usepackage{graphicx}
\usepackage{dcolumn}
\usepackage{bm}

\begin{document}

\title{{\em Zitterbewegung} of electrons and holes in III-V
semiconductor quantum wells}

\author{John Schliemann$^{1}$, Daniel Loss$^{2}$, and R.~M. Westervelt$^{3}$}

\address{$^{1}$Institute for Theoretical Physics, University of Regensburg, 
D-93040 Regensburg, Germany\\
$^{2}$Department of Physics and Astronomy, University of Basel,
CH-4056 Basel, Switzerland\\
$^{3}$Division of Engineering and Applied Sciences, Harvard University, 
Cambridge, Massachusetts 02138, USA}

\date{\today}

\begin{abstract}
The notion of {\em zitterbewegung} is a long-standing prediction of 
relativistic quantum mechanics. Here we extend earlier theoretical studies on
this phenomenon for the case of III-V zinc-blende semiconductors which exhibit 
particularly strong spin-orbit coupling. This property makes nanostructures
made of these materials very favorable systems for possible experimental
observations of {\em zitterbewegung}. Our investigations include electrons
in n-doped quantum wells under the influence of both Rashba and Dresselhaus
spin-orbit interaction, and also the two-dimensional hole gas. Moreover, we 
give a detailed anaysis of electron {\em zitterbewegung} in quantum wires
which appear to be particularly suited for experimentally observing this 
effect.
\end{abstract}
\vskip2pc]

\maketitle

\section{introduction}
\label{intro}

The spin degree of freedom of electrons in semiconductor nanostructures
is one of the central subjects in the growing field of spin electronics
\cite{overview}. The latter key word describes the whole variety of 
efforts and 
proposals for using the electron spin instead, or in combination with, its
charge for information processing, or, even more ambitious, quantum
information processing. A central issue is the possibility of {\em electrical}
control electron spins, which avoids many difficulties arising from applying
and gating magnetic fields.
Zinc-blende III-V semiconductors show a particularly
strong spin-orbit interaction and are therefore natural candidates with
respect to the above goals.

In the present paper we extend earlier results on the relativistic effect of 
{\em zitterbewegung} in systems of the above type \cite{Schliemann05a}.
In this context we also discuss related recent work by other authors
\cite{Jiang05,Zawadzki05a,Lee05,Nikolic05,Shen05,Zawadzki05b}. Our 
investigations include electrons
in n-doped quantum wells under the influence of both Rashba and Dresselhaus
spin-orbit interaction, and holes in two-dimensional geometry.
Phenomena related to spin-orbit coupling in such systems are presently
studied very intensively also in the context of the intrinsic spin-Hall effect
\cite{Murakami03,Sinova04,Schliemann05b}. Particular attention is paid to the 
case of electron {\em zitterbewegung} in quantum wires
which appear to be good candidates for experimental investigations of this 
effect.

The coupling between the orbital and the spin degree of freedom of electrons 
is a relativistic effect described by the Dirac equation and its 
nonrelativistic
expansion in powers of the inverse speed of light $c$ \cite{Feshbach58}. 
In second order
one obtains, apart from two spin-independent contributions, the following 
well-known spin-orbit coupling term,
\begin{equation}
{\cal H}_{so}=\frac{1}{2m_{0}c^{2}}\vec s\cdot\left(\nabla V\times\frac{\vec p}{m_{0}}\right)\,,
\label{sogeneral}
\end{equation}
where $m_{0}$ is the bare mass of the electron, $\vec s$, $\vec p$ its spin and
momentum, respectively, and $V$ is some applied external potential. On the 
other hand, the free Dirac equation, $V=0$, has two dispersion branches with
positive and negative energy, 
\begin{equation}
\varepsilon(\vec p)=\pm\sqrt{m_{0}^{2}c^{4}+c^{2}p^{2}}\,,
\end{equation}
which are separated by an energy gap of $2m_{0}c^{2}\approx 1{\rm MeV}$. In particular,
the nonrelativistic expansion of the Dirac equation quoted above can be seen
as a method of systematically including the effects of the negative-energy
solutions on the states of positive energy starting from their nonrelativistic
limit \cite{Feshbach58}. Moreover, the large energy gap $2m_{0}c^{2}$ appears in the
denominator of the right hand side of Eq.~(\ref{sogeneral}), suppressing the
effects of spin-orbit coupling for weakly bound electrons.
 
On the other hand, the band structure of zinc-blende III-V semiconductors
shows many formal similarities to the situation of free relativistic electrons,
while the relevant energy scales are grossly different
\cite{Zawadzki70,Darnhofer93,Rashba04}. For not too large
doping of such semiconductors, one can concentrate on the band structure around
the $\Gamma$ point. Here one has a parabolic $s$-type conduction band and a
$p$-type valence band consisting of the well-known dispersion branches
for heavy and light holes, and the split-off band. However, the gap between 
conduction and valence band is of order $1{\rm eV}$ or smaller. This heuristic
argument makes plausible that spin-orbit coupling is an important effect in
III-V semiconductors which actually lies at the very heart of the field of
semiconductor spintronics.

Dating back to a seminal paper by Schr\"odinger \cite{Schrodinger30,Barut81}
from 1930, the notion of {\em zitterbewegung} has been a long-standing 
theoretical
prediction of relativistic quantum mechanics. In the free Dirac equation, this
oscillatory quantum motion occurs for particle wave packets which are 
superpositions of both solutions of positive and negative energy. Thus, the
dominant frequency of this dynamics is given by $2m_{0}c^{2}/{\hbar}$ which is of order
$10^{20}{\rm Hz}$. Moreover, the length scale of this motion is given by
the Compton wave length $\hbar/m_{0}c$ of the free electron. Therefore, in order to
experimentally observe the {\em zitterbewegung} of free electrons one would
need to confine these objects on a length scale of a few picometers. Now it
follows form general uncertainty arguments that such a spatial confinement
leads to an energy scale where electron-positron pair production plays a
serious and detrimental role. These arguments have led many authors to the 
opinion that the {\em zitterbewegung} of electrons is impossible to observe, 
see e.g. Ref.~\cite{Huang52}.

Most recently, we have theoretically investigated {\em zitterbewegung} in
III-V zinc-blende semiconductors and developed a proposal for its experimental
observation. As already mentioned, the mathematical treatment of effective
band structure models relevant to such systems on the one hand, and the
nonrelativistic expansion of the Dirac equation on the other hand show
many formal similarities \cite{Zawadzki70,Darnhofer93,Rashba04}. In fact, 
under certain aspects, the
low-energy band structure of such semiconductors around the Fermi level can be
viewed as a model for truly relativistic electrons, but with energy and
length scales which are much more favourable for observing effects like
{\em zitterbewegung}.

This paper is organized as follows. In section \ref{2DEG} we study 
{\em zitterbewegung} in the two-dimensional electron gas. We extend previous
results \cite{Schliemann05a} to the situation where spin-orbit
coupling of both the Rashba and Dresselhaus type is present. Moreover, we
give a deeper discussion of {\em zitterbewegung} in quantum wires and our
related experimental proposal. 
Section \ref{2Dholes} is devoted to the same phenomenon for heavy holes
in two-dimensional quantum wells. We close with conclusions in section
\ref{conclusions}.

\section{{\em Zitterbewegung} in the two-dimensional electron gas 
and in quantum wires} 
\label{2DEG}

\subsection{The two-dimensional electron gas}

For conduction band electrons in $n$-doped quantum wells of 
zinc-blende semiconductor  structures the dominant
effects of spin-orbit interaction 
can be described in terms of two effective contributions to the Hamiltonian. 
On of them is the Rashba spin-orbit term \cite{Rashba60}
which is due to the inversion-asymmetry of the
confining potential and has the form
\begin{equation}
{\cal
H}_{R}=\frac{\alpha}{\hbar}\left(p_{x}\sigma^{y}-p_{y}\sigma^{x}\right)\,,
\label{rashba}
\end{equation}
where $\vec p$ is the momentum of the electron confined in a
two-dimensional geometry, and $\vec\sigma$ the vector of Pauli
matrices. The coefficient $\alpha$ is tunable in strength by the
external gate perpendicular to the plane of the two-dimensional electron gas.
The other contribution is the Dresselhaus spin-orbit term which is present
in semiconductors lacking bulk inversion symmetry\cite{Dresselhaus55}. 
When restricted to a two-dimensional semiconductor nanostructure grown along 
the $[001]$ direction this coupling is of the form 
\cite{Dyakonov86,Bastard92}
\begin{equation}
{\cal H}_{D}=\frac{\beta}{\hbar}\left(p_{y}\sigma^{y}-p_{x}\sigma^{x}\right)\,,
\label{dressel}
\end{equation}
where the coefficient $\beta$ is determined by the semiconductor
material and the geometry of the sample. These two contributions to
the effective Hamiltonian have also an interesting interplay
\cite{Schliemann03a,Schliemann03b}.

We now consider the single-particle Hamiltonian of a free electron under the 
influence of spin-orbit coupling of both the Rashba and the Dresselhaus type,
\begin{equation}
{\cal H}=\frac{\vec p^{2}}{2m}+{\cal H}_{R}+{\cal H}_{D}\,,
\end{equation}
where $m$ is the effective band mass. The components of the time-dependent
position operator
\begin{equation}
\vec r_{H}(t)=e^{i{\cal H}t/\hbar}\vec r(0)e^{-i{\cal H}t/\hbar}
\end{equation}
in the Heisenberg picture read explicitly
\begin{eqnarray}
x_{H}(t) & = & x(0)+ \frac{p_{x}}{m}t
+\left(\frac{\alpha}{\hbar}\sigma^{y}-\frac{\beta}{\hbar}\sigma^{x}\right)t\nonumber\\
 & + & \frac{\alpha^{2}-\beta^{2}}{\Lambda^{2}}p_{y}
\left(1-\cos\left(\frac{2}{\hbar^{2}}\Lambda t\right)\right)\frac{\hbar}{2}\sigma^{z}\nonumber\\
 & + & \frac{\alpha^{2}-\beta^{2}}{\Lambda^{3}}p_{y}
\left(\frac{2}{\hbar^{2}}\Lambda t
-\sin\left(\frac{2}{\hbar^{2}}\Lambda t\right)\right)\Sigma \,,
\label{xHgen}\\
y_{H}(t) & = & x(0)+ \frac{p_{y}}{m}t
-\left(\frac{\alpha}{\hbar}\sigma^{x}-\frac{\beta}{\hbar}\sigma^{y}\right)t\nonumber\\
 & - & \frac{\alpha^{2}-\beta^{2}}{\Lambda^{2}}p_{x}
\left(1-\cos\left(\frac{2}{\hbar^{2}}\Lambda t\right)\right)\frac{\hbar}{2}\sigma^{z}\nonumber\\
 & - & \frac{\alpha^{2}-\beta^{2}}{\Lambda^{3}}p_{x}
\left(\frac{2}{\hbar^{2}}\Lambda t
-\sin\left(\frac{2}{\hbar^{2}}\Lambda t\right)\right)\Sigma
\label{yHgen}
\end{eqnarray}
with
\begin{equation}
\Lambda^{2}(\vec p;\alpha,\beta)=\left(\alpha^{2}+\beta^{2}\right)p^{2}+4\alpha\beta p_{x}p_{y}
\end{equation}
and
\begin{eqnarray}
\Sigma(\vec p,\vec\sigma ;\alpha,\beta) & = & \frac{\hbar}{2}\Bigl(
\alpha\left(p_{x}\sigma^x+p_{y}\sigma^y\right)\nonumber\\
 & & +\beta\left(p_{x}\sigma^y+p_{y}\sigma^x\right)\Bigr)\,.
\end{eqnarray}
Here the operators $\vec p$ and $\vec\sigma$ on the right hand sides are
in the Schr\"odinger picture and therefore time-independent.

The oscillatory terms on the right hand sides of Eqs.~(\ref{xHgen}),
(\ref{yHgen}) can be viewed as the {\em zitterbewegung} the electron performs 
under the influence of spin-orbit coupling. This oscillatory quantum motion
vanishes if relativistic effects are absent, $\alpha=\beta=0$. The contributions linear 
in time $t$ in the first lines in Eqs.~(\ref{xHgen}),(\ref{yHgen}) are just
proportional to the velocity 
\begin{equation}
\vec v=\frac{i}{\hbar}\left[{\cal H},\vec r\right]\,,
\end{equation}
which is in the presence of spin-orbit coupling a spin-dependent operator.
The expressions (\ref{xHgen}),(\ref{yHgen}) contain the results given in Ref.
\cite{Schliemann05a} for pure Rashba or Dresselhaus coupling as special cases.
For, instance, if only Rashba coupling is present $(\beta=0)$, one can express the
position operator in the Heisenberg picture as
\begin{eqnarray}
x_{H}(t) & = & x(0)+ \frac{p_{x}}{m}t
+\frac{p_{y}}{p^{2}}\frac{\hbar}{2}\sigma^{z}
\left(1-\cos\left(\frac{2\alpha p}{\hbar^{2}}t\right)\right)\nonumber\\
 & + & \frac{p_{x}}{p^{3}}\frac{\hbar}{2}
\left(p_{x}\sigma^{y}-p_{y}\sigma^{x}\right)
\left(\frac{2\alpha p}{\hbar^{2}}t
-\sin\left(\frac{2\alpha p}{\hbar^{2}}t\right)\right)\nonumber\\
 & + & \frac{1}{p}\frac{\hbar}{2}\sigma^{y}
\sin\left(\frac{2\alpha p}{\hbar^{2}}t\right)\,,
\label{xHrashba}\\
y_{H}(t) & = & y(0)+ \frac{p_{y}}{m}t
-\frac{p_{x}}{p^{2}}\frac{\hbar}{2}\sigma^{z}
\left(1-\cos\left(\frac{2\alpha p}{\hbar^{2}}t\right)\right)\nonumber\\
 & + & \frac{p_{y}}{p^{3}}\frac{\hbar}{2}
\left(p_{x}\sigma^{y}-p_{y}\sigma^{x}\right)
\left(\frac{2\alpha p}{\hbar^{2}}t
-\sin\left(\frac{2\alpha p}{\hbar^{2}}t\right)\right)\nonumber\\
 & - & \frac{1}{p}\frac{\hbar}{2}\sigma^{x}
\sin\left(\frac{2\alpha p}{\hbar^{2}}t\right)\,.
\label{yHrashba}
\end{eqnarray}
The case $\alpha=\pm\beta$ is particular \cite{Schliemann03a,Schliemann03b}. As seen from
Eqs.~(\ref{xHgen}),(\ref{yHgen}), the oscillatory part of the time-dependent
position operator, i.e. the {\em zitterbewegung} vanishes due to the
prefactor $(\alpha^{2}-\beta^{2})$. Physically, this observation results from an additional
conserved quantity arising at this point \cite{Schliemann03a,Schliemann03b}.

It is straightforward to evaluate the above time-dependent position operators
within Gaussian wave packets. For simplicity we concentrate on the case 
of pure Rashba coupling described by Eqs~(\ref{xHrashba}),(\ref{yHrashba}).
We consider a Gaussian wave packet with initial spin polarization along the 
$z$-direction perpendicular to the quantum well,
\begin{equation}
\langle\vec r|\psi\rangle=\frac{1}{2\pi}\frac{d}{\sqrt{\pi}}
\int d^{2}k\,e^{-\frac{1}{2}d^{2}\left(\vec k-\vec k_{0}\right)^{2}}
e^{i\vec k\vec r}
\left(
\begin{array}{c}
1 \\ 0
\end{array}
\right)\,.
\label{initz}
\end{equation}
Clearly we have $\langle\psi|\vec r|\psi\rangle=0$,
$\langle\psi|\vec p|\psi\rangle=\hbar\vec k_{0}$, and the variances 
of the position and momentum operators are 
$\left(\Delta x\right)^{2}=\left(\Delta y\right)^{2}=d^{2}/2$,
$\left(\Delta p_{x}\right)^{2}=\left(\Delta p_{y}\right)^{2}
=\hbar^{2}/2d^{2}$.
Thus, the group velocity of the wave packet is given by 
$\hbar\vec k_{0}/m$ , while 
its spatial width is described by the parameter $d$ with the minimum
uncertainty product typical for Gaussian wave packets,
$\Delta p_{x}\Delta x=\Delta p_{y}\Delta y=\hbar/2$.

A direct calculation gives
\begin{eqnarray}
\langle\psi|x_{H}(t)|\psi\rangle & = & \frac{\hbar k_{0x}}{m}
+\frac{d}{2\pi}e^{-d^{2}k_{0}^{2}}\int_{0}^{2\pi}d\varphi\sin\varphi
\nonumber\\
& & \cdot\int_{0}^{\infty}dq\,e^{-q^{2}+2dq
\left(k_{0x}\cos\varphi+k_{0y}\sin\varphi\right)}\nonumber\\
& & \qquad\cdot\left(1-\cos\left(\frac{2\alpha q}{\hbar d}t\right)\right)\,.
\label{xHintegral}
\end{eqnarray}
In the above expression $q$ is a dimensionless integration variable.
The remaining integration over the polar angle $\varphi$ gives a vanishing result
if $k_{0y}=0$, i.e. if the group velocity
is along the $x$-direction. More generally, one finds that
\begin{equation}
\langle\psi|\vec k_{0}\cdot\vec r_{H}(t)/k_{0}|\psi\rangle
=\frac{\hbar k_{0}}{m}t\,,
\end{equation}
which means that the {\em zitterbewegung} is always perpendicular to the group
velocity of the wave packet. The same observation is made for the case
of pure Dresselhaus coupling, $\alpha=0$. Note that for pure Rashba coupling the
sum of the $z$-component of the orbital angular momentum 
$\vec l=\vec r\times\vec p$ and the $z$-component of the spin $\vec s=\hbar\vec\sigma/2$ is
a conserved quantity, $[{\cal H},l^{z}+s^{z}]=0$, while for pure Dresselhaus
coupling we have $[{\cal H},l^{z}-s^{z}]=0$. In the presence of these conservation
laws, the {\em zitterbewegung} can also be interpreted as a consequence of spin
rotation due to spin-orbit coupling. Consider for instance an electron moving 
along the $y$-direction with its spin being initially aligned with the
$z$-direction. In the time evolution of the particle, the spin will then be 
rotated due spin-orbit coupling which requires, by virtue of the conservation 
law, also a finite component $l^{z}$ to develop, i.e. the electron has to
perform also a movement perpendicular to its group velocity. In the
general case $\alpha\neq 0\neq\beta$ such an interpretation does not seem to be possible since
a conserved quantity of the above kind does not exist.

Let us turn back to the case of pure Rashba coupling. Without loss of
generality we consider an electron wave packet moving along the $y$-direction,
$k_{0x}=0$. By expanding the exponential containing the trigonometric functions
in Eq.~(\ref{xHintegral}), one derives \cite{noteerratum}
\begin{eqnarray}
\langle\psi|x_{H}(t)|\psi\rangle  & = & \frac{1}{2k_{0y}}
\left(1-e^{-d^{2}k_{0y}^{2}}\right)\nonumber\\
 & -  & \frac{1}{k_{0y}}e^{-d^{2}k_{0y}^{2}}\sum_{n=0}^{\infty}\Biggl[
\frac{\left(dk_{0y}\right)^{2(n+1)}}{n!(n+1)!}\nonumber\\
 & & \cdot\int_{0}^{\infty}dq q^{2n+1}e^{-q^{2}}
\cos\left(\frac{2\alpha q}{\hbar d}t\right)\Biggr]\,.
\label{xHexpansion}
\end{eqnarray}
Thus, the amplitude of the {\em zitterbewegung} is proportional to the
wave length $\lambda_{0y}=2\pi/k_{0y}$ of the electron motion perpendicular to it. In a 
semiconductor quantum well, this length can be of order a few ten nanometers, 
which is several orders of magnitude larger than the length scale of
the {\em zitterbewegung} of free electrons given by the Compton wave length.
Note also that the oscillatory
{\em zitterbewegung} changes its sign if the translational motion is reversed.

If the product $dk_{0y}$ is not too large, $dk_{0y}\lesssim 1$, only low
values of the summation index $n$ in Eq.~(\ref{xHexpansion})
lead to substantial contributions, and the
Gaussian factor in the integrand suppresses contributions from large values
of $q$. Thus, a typical scale of this integration variable is leading to
sizable contributions is $q\approx 1/\sqrt{2}$. Thus, a typical time scale
in the integrand is $T=\sqrt{2}\pi\hbar d/\alpha$, and when averaging the
zitterbewegung over times scales significantly larger than $T$, the
cosine term drops giving
\begin{equation}
\overline{\langle\psi|x_{H}(t)|\psi\rangle}
=\frac{1}{2k_{0y}}\left(1-e^{-d^{2}k_{0y}^{2}}\right)\,,
\end{equation} 
i.e. the time-averaged guiding center of the wave packet is shifted 
perpendicular to its direction of motion. Note that the zitterbewegung is
absent for $k_{0y}=0$ \cite{Huang52}.

In the opposite case $dk_{0y}\gg 1$ the Gaussian approaches a 
$\delta$-function. In this limit one finds (for $k_{0x}=0$)
\begin{equation}
\langle\psi|x_{H}(t)|\psi\rangle=
\frac{1}{2k_{0y}}\left(1
-\cos\left(\frac{2\alpha k_{0y}}{\hbar}t\right)\right)\,.
\end{equation}
Here the frequency of the zitterbewegung is $\Omega=2\alpha k_{0y}/\hbar$, 
and the guiding center of the wave packet is also
shifted in the direction perpendicular to its group velocity. Note
that $\hbar\Omega$ is the excitation energy between the two branches of the
Rashba Hamiltonian ${\cal H}$ at a given momentum $\vec k=k_{0y}\vec e_{y}$.
Rashba spin-orbit coupling is particularly strong in InAs where values for the
parameter $\alpha$ of a few $10^{-11}{\rm eVm}$ can be reached
\cite{Nitta97,Engels97,Heida98,Hu99,Grundler00,Sato01,Hu03}, leading to
frequencies $\Omega$ is the terahertz regime. This is much smaller than
the typical frequency of the {\em zitterbewegung} of free electrons which is
of order $10^{20}{\rm Hz}$. For GaAs, the Rashba coefficient is typically an 
order of magnitude smaller than in InAs \cite{Miller03} and the Dresselhaus
coupling plays a more important role \cite{Lommer88,Jusserand92,Jusserand95}.
In summary, the {\em zitterbewegung} of electronic wave packets in 
semiconductor
quantum wells as discussed above is characterized by amplitudes and
frequencies which are by orders of magnitude larger and smaller, 
respectively, than it is the case for free electrons. This opens the 
perspective to experimentally observe the electron {\em zitterbewegung}
in semiconductor nanostructures via terahertz methods, or
using high-resolution scanning-probe 
microscopy imaging techniques as developed and discussed in 
Refs.~\cite{Topinka00,LeRoy03}. The latter approach will be described in more
detail in section \ref{wires}.

The issue of {\em zitterbewegung} of electrons in III-V semiconductors
was also discussed recently by Zawadzki where the three-dimensional bulk case
was considered \cite{Zawadzki05a}. The author starts from an $8\times8$ Kane model
for conduction and valence band being diagonal in the hole sector
\cite{Darnhofer93,Vurgaftman01}.
Specializing on particles moving along the $z$-direction in real space
and neglecting the split-off band, the author obtains an effective Hamiltonian
coupling only light holes and conduction band electron states. Moreover, for an
appropriate choice of basis this Hamiltonian matrix mimics the Hamiltonian of
the free Dirac equation, which enables to derive a {\em zitterbewegung}
following directly Schr\"odinger's original approach 
\cite{Schrodinger30,Barut81}.
However, when reducing the underlying $8\times8$ Kane model in a systematic way
to an effective $2\times2$ Hamiltonian for conduction band electrons only, one
obtains in the absence of an additional potential and magnetic fields
to second order in the gap energy only a kinetic term involving an
effective mass depending on band structure parameters \cite{Darnhofer93}. This
is analogous to the situation of the free Dirac equation where, again
in the absence of external fields, the lowest-order relativistic correction
is an additional contribution to the kinetic energy which does not lead
to {\em zitterbewegung} \cite{Feshbach58}. 
The {\em zitterbewegung} studied in the present work
occurs, for the case of Rashba coupling, due to an external potential
introducing structure-inversion asymmetry \cite{Rashba60,Darnhofer93}.
For the case of Dresselhaus coupling, it stems from the bulk Dresselhaus
coupling term which results from bulk-inversion asymmetry 
\cite{Dresselhaus55} and is not included in the $8\times8$ Kane model.
In this sense, Zawadzki's result appears to be
effect of higher order in the inverse gap energy which can be of importance
in materials with particularly small gap such as InSb \cite{Vurgaftman01}.
Moreover, the {\em zitterbewegung} as discussed in Ref.~\cite{Zawadzki05a}
occurs always in in the direction of the group velocity of the
particle wave packet, i.e. in the $z$-direction. This feature is clearly
different from the {\em zitterbewegung} investigated here and might pose
an obstacle against experimentally observing this effect. 

\subsection{Quantum wires}
\label{wires}

The {\em zitterbewegung} of an electron in a quantum well as described above
is naturally accompanied by a broadening of the wave packet, where the
dominant contribution stems from the dispersive effective-mass term in the
Hamiltonian. Such a broadening might pose an obstacle for experimentally
detecting the zitterbewegung. However, the broadening can be efficiently 
suppressed and limited if the electron moves along a quantum wire.
In fact, the motion of electrons in quantum wells is generally under better 
control if additional lateral confinement is present. 
We therefore consider a harmonic
quantum wire along the $y$-direction described by
\begin{equation}
{\cal H}=\frac{\vec p^{2}}{2m}+\frac{1}{2}m\omega^{2}x^{2}+{\cal H}_{R}\,, 
\end{equation}
where the frequency
$\omega$ parameterizes the confining potential perpendicular to the wire
\cite{wire,Governale02}. It is instructive to write the Hamiltonian in the form
\begin{equation}
{\cal H}={\cal H}_{0}+{\cal H}_{1}
\end{equation}
with
\begin{eqnarray}
{\cal H}_{0} & = & \hbar\omega\left(a^{+}a+\frac{1}{2}\right)
+\frac{\hbar^{2}k_{0y}^{2}}{2m}+\alpha k_{0y}\sigma^{x}\,,\\
{\cal H}_{1} 
& = & -i\sqrt{\frac{\hbar m\omega}{2}}\frac{\alpha}{\hbar}\left(a-a^{+}\right)\sigma^{y}\,.
\end{eqnarray}
Here $a$, $a^{+}$ are the usual harmonic climbing operators, and $k_{0y}$ 
is the 
component of the electron wave vector along the quantum wire. 
Due to the properties of the 
``mixing operator'' ${\cal H}_{1}$, analytical progress as before 
without employing further approximations does not
seem to be possible. We therefore project the Hamiltonian ${\cal H}$ onto
the lowest two orbital subbands. This approximation is known to give very
reasonable results for not too wide quantum wells \cite{Governale02}, and we 
will also compare its results with a full numerical simulation of the above
multi-band Hamiltonian.

\begin{figure}
\centerline{\includegraphics[width=8cm]{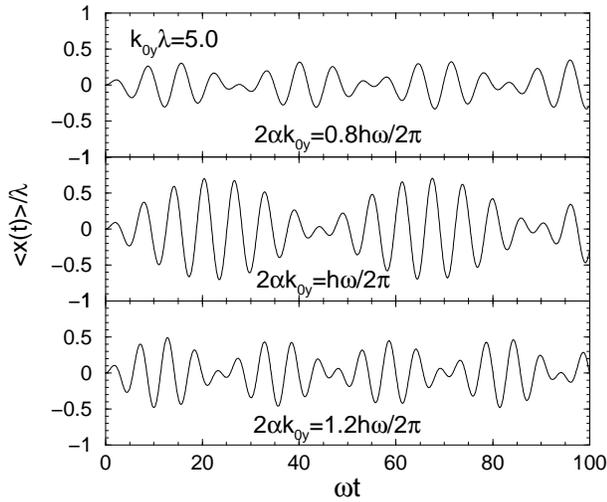}} 
\caption{{\em Zitterbewegung} of an electron in a harmonic quantum wire
perpendicular to the wire direction. The electron is initially injected
into the lowest subband with its spin pointing along the $z$-direction.
The wave number $k_{0y}$ for the motion 
along the wire is $k_{0y}\lambda=5$. The amplitude of the {\em zitterbewegung} 
is maximal at the resonance $2\alpha k_{0y}=\hbar\omega$
(middle panel).
\label{fig1}}
\end{figure}

\begin{figure}
\centerline{\includegraphics[width=8cm]{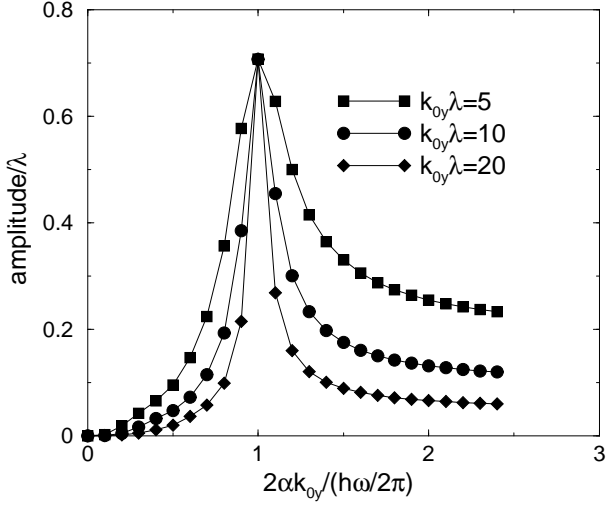}} 
\caption{Amplitude of the {\em zitterbewegung} perpendicular to the wire 
direction
as a function of $\Omega/\omega=2\alpha k_{0y}/\hbar\omega$ for different 
values of the wave number $k_{0y}$ along the wire. Again the electron spin points
initially in the $z$-direction.
\label{fig2}}
\end{figure}

For a given $k_{0y}$ the truncated
Hilbert space is spanned by the states $|0,\uparrow\rangle$, 
$|0,\downarrow\rangle$, $|1,\uparrow\rangle$, $|1,\downarrow\rangle$, 
where the arrows denote the spin state with respect to the $z$-direction,
and $0$ and $1$ stand for the ground state and the first excited state of
the harmonic potential, respectively. In the above basis, the 
truncated Hamiltonian reads
\begin{equation}
{\cal H}=\left(
\begin{array}{cccc}
\varepsilon_{0} & -\alpha k_{0y} & 0 & -\sqrt{\frac{\hbar\omega\varepsilon_{R}}{2}} \\
-\alpha k_{0y} & \varepsilon_{0} & \sqrt{\frac{\hbar\omega\varepsilon_{R}}{2}} & 0 \\
0 & \sqrt{\frac{\hbar\omega\varepsilon_{R}}{2}}  & \varepsilon_{1} & -\alpha k_{0y} \\
-\sqrt{\frac{\hbar\omega\varepsilon_{R}}{2}}  & 0 & -\alpha k_{0y} & \varepsilon_{1} 
\end{array}
\right)
\end{equation}
with
\begin{eqnarray}
\varepsilon_{0} & = & \frac{1}{2}\hbar\omega+\frac{\hbar^{2}k_{0y}^{2}}{2m}\,,\\
\varepsilon_{1} & = & \frac{3}{2}\hbar\omega+\frac{\hbar^{2}k_{0y}^{2}}{2m}
\end{eqnarray}
being the subband energies in the absence of spin-orbit coupling, and 
$\varepsilon_{R}=m\alpha^{2}/\hbar^{2}$ is the energy scale of the Rashba coupling.
When applying the transformation
\begin{equation}
U=\frac{1}{\sqrt{2}}\left(
\begin{array}{cccc}
1 & 1 & 0 & 0  \\
0 & 0 & 1 & -1 \\
1 & -1 & 0 & 0 \\
0 & 0 & 1 & 1 
\end{array}
\right)
\end{equation}
the projected Hamiltonian and in turn the time evolution operator become
block-diagonal,
\begin{equation}
Ue^{-\frac{i}{\hbar}Ht}U^{+}=\left(
\begin{array}{cc}
h_{+}(t) & 0 \\
0 & h_{-}(t)
\end{array}
\right)
\end{equation}
where
\begin{eqnarray}
h_{\pm}(t) & = & 
\left({\bf 1}\cos\left(\frac{1}{\hbar}\mu_{\pm}t\right)
-i\frac{\vec\mu_{\pm}}{\mu_{\pm}}\cdot\vec\sigma
\sin\left(\frac{1}{\hbar}\mu_{\pm}t\right)\right)\nonumber\\
 & & \cdot\exp\left(-\frac{i}{\hbar}\left(\hbar\omega
+\frac{\hbar^{2}k_{0y}^{2}}{2m}\right)t\right)
\end{eqnarray}
and 
\begin{equation}
\vec\mu_{\pm}=\left(\pm\alpha\sqrt{\frac{m\omega}{2\hbar}},0,-\frac{\hbar\omega}{2}\mp\alpha k_{0y}\right)\,.
\end{equation}
Let us first consider an electron with a given momentum $k_{0y}$ along the wire 
and injected initially into the lowest subband of the confining potential
with the spin pointing upwards 
along the $z$-direction, i.e. the initial wave function $|\psi_{z}\rangle=|0,\uparrow\rangle$
for the $x$-direction
is a Gaussian whose width is determined by the characteristic length
$\lambda=\sqrt{\hbar/m\omega}$ of the harmonic confinement. Using 
\begin{equation}
UxU^{+}=\frac{\lambda}{\sqrt{2}}\left(
\begin{array}{cc}
0 & \sigma^{x} \\
\sigma^{x} & 0 
\end{array}
\right)
\end{equation}
one obtains for the above initial state the
following time-dependent expectation value
\begin{eqnarray}
\langle\psi_{z}|x_{H}(t)|\psi_{z}\rangle & = &
\frac{\lambda}{\sqrt{2}}\frac{\mu_{+}^{x}\mu_{-}^{z}+\mu_{-}^{x}\mu_{+}^{z}}
{\mu_{+}\mu_{-}}\nonumber\\
 & & \cdot\sin\left(\frac{1}{\hbar}\mu_{+}t\right)
\sin\left(\frac{1}{\hbar}\mu_{-}t\right)\,.
\end{eqnarray}
The amplitude of this oscillatory dynamics perpendicular to the wire direction
becomes maximal when the resonance condition $\hbar\omega=\pm2\alpha k_{0y}$ is fulfilled. At
that point we have $\mu_{\mp}^{z}=0$,
and if $\mu_{\pm}^{x}$ can be neglected compared to $\mu_{\pm}^{z}$
(which is the case for large enough $k_{0y}$) the amplitude of the 
zitterbewegung is approximately $\lambda/\sqrt{2}$. This result from the
truncated Hamiltonian is in excellent agreement with numerical simulations
of the full multi-band system. 
In Fig.~\ref{fig1} we have plotted simulation results for 
$\langle\psi_{z}|x_{H}(t)|\psi_{z}\rangle$ where the wave number
along the wire is fixed to be $k_{0y}\lambda=5$ and the Rashba parameter
$\alpha$ is varied around the resonance condition. Clearly, the amplitude is
maximum at resonance.Equivalent observation are 
made if the Rashba coupling is fixed while the wave number $k_{0y}$ is varied.
In Fig.~\ref{fig2} we have plotted the amplitude of the zitterbewegung
as a function of $\Omega/\omega=2\alpha k_{0y}/\hbar\omega$ for different 
values of the wave number $k_{0y}$ along the wire. In this range of parameters,
the resonance becomes narrower with increasing $k_{0y}$, while
its maximum value is rather independent of this quantity and 
remarkably well described by $\lambda/\sqrt{2}$. 

A qualitative explanation
for this resonance can be given in terms of the decomposition
${\cal H}={\cal H}_{0}+{\cal H}_{1}$ of the Hamiltonian.  The 
{\em zitterbewegung} is induced by the perturbation ${\cal H}_{1}$ which can 
act most efficiently if the unperturbed energy levels of ${\cal H}_{0}$ are 
degenerate having opposite spins. This is the case at 
$|2\alpha k_{0y}|=\hbar\omega$.

We propose that electron zitterbewegung in semiconductor nanostructures
as described above can be experimentally observed using high-resolution
scanning-probe microscopy 
imaging techniques as developed and discussed in 
Refs.~\cite{Topinka00,LeRoy03}. As a possible setup, a tip 
can be moved along the wire and 
centered in its middle. For an appropriate biasing of the tip, the 
electron density at its location is partially
depleted leading to a reduced conductance
of the wire. Since the amplitude of the
zitterbewegung reflects the electron density in the center of the wire, 
the zitterbewegung will induce beatings in the wire conductance as
a function of the tip position. These beatings are most pronounced at the
resonance, see Fig.~\ref{fig1}. Note that the oscillations shown there as
a function of time can be easily converted to the real-space $y$-coordinate
by multiplying the abscissa by $\hbar k_{0y}/m$. Moreover, for tuning the system to
the resonance condition $|2\alpha k_{0y}|=\hbar\omega$, at least two 
parameters can be varied 
expertimentally: The group velocity of the injected electron along the wire
given by $k_{0y}$, and the Rashba parameter $\alpha$ which is tunable by a gate
voltage across the quantum well. Thus, quantum wires defined in InAs quantum 
wells are favorable systems for experiments of the above kind, since this 
material can exhibit a quite large Rashba coupling 
but has a comparatively small
Dresselhaus term. The group velocity can be varied by changing the 
gate to the two-dimensional electron gas. This alters the global electron
density and therefore also the wave vector for motion along the wire.
Generally we expect spin-orbit
effects in STM experiments to be more pronounced in the presence of
additional confinement such as in a quantum wire. 

The present considerations
concentrate on the case of pure Rashba coupling neglecting a possible
Dresselhaus contribution. Very analogous observations as above can be made for
pure Dresselhaus coupling, while in the case of both couplings being present
the analytical theory becomes technically more involved. We note that the
Dresselhaus coupling, differently from the Rashba term, cannot be tuned by an 
external gate. Therefore, materials with pronounced Rashba coupling are
favorable for tuning the system to the above resonance condition.

Let us now analyze the situation when the electron is injected into the 
lowest subband of the wire, but its spin is not aligned with the $z$-direction.
If the spin points along the $x$-direction, no {\em zitterbewegung}
occurs,
\begin{equation}
\langle\psi_{x}|x_{H}(t)|\psi_{x}\rangle=0\,.
\end{equation}
This is a property of both the full multi-band model and the truncated 
Hamiltonian and follows from symmetry considerations: Under a reflection in
the $yz$-plane $x$, $p_{x}$ and the $y$- and $z$-component of the spin 
change sign while
the other components of spin and momentum remain unchanged. Thus, the 
Hamiltonian is invariant under this operation, a property which is also 
shared by the initial state $|\psi_{x}\rangle$. Therefore, the expectation value of
$x_{H}(t)$ has to be equal to its negative and is consequently zero. Finally, if
the spin points initially along the $y$-direction one finds from the
truncated Hamiltonian
\begin{eqnarray}
\langle\psi_{y}|x_{H}(t)|\psi_{y}\rangle & = &
\frac{\lambda}{\sqrt{2}}\frac{1}{2}\Biggl[
\frac{\mu_{+}^{x}\mu_{-}^{z}+\mu_{-}^{x}\mu_{+}^{z}}
{\mu_{+}\mu_{-}}\nonumber\\
 & & \cdot\sin\left(\frac{1}{\hbar}\mu_{+}t\right)
\sin\left(\frac{1}{\hbar}\mu_{-}t\right)\nonumber\\
 & + & \frac{\mu_{+}^{x}}{\mu_{+}}\cos\left(\frac{1}{\hbar}\mu_{+}t\right)
\sin\left(\frac{1}{\hbar}\mu_{-}t\right)\nonumber\\
 & - & \frac{\mu_{-}^{x}}{\mu_{-}}\sin\left(\frac{1}{\hbar}\mu_{+}t\right)
\cos\left(\frac{1}{\hbar}\mu_{-}t\right)\Biggr]\,.
\end{eqnarray}
Thus, the {\em zitterbewegung} also occurs if the electron spin is initially
aligned along the wire direction.

It is instructive to also investigate the dynamics of the spin degree of 
freedom as the electron passes along the wire. For a situation where 
the spin is pointing again 
initially in the $z$-direction, the truncated Hamiltonian leads 
to the following expressions:
\begin{eqnarray}
& & \langle\psi_{z}|\sigma^{x}_{H}(t)|\psi_{z}\rangle=\frac{1}{2}\Biggl[
\cos^{2}\left(\frac{1}{\hbar}\mu_{+}t\right)\nonumber\\
 &  & \qquad+\frac{-\left(\mu_{+}^{x}\right)^{2}+\left(\mu_{+}^{z}\right)^{2}}{\mu_{+}^{2}}
\sin^{2}\left(\frac{1}{\hbar}\mu_{+}t\right)\nonumber\\
 & & \qquad-\cos^{2}\left(\frac{1}{\hbar}\mu_{-}t\right)\nonumber\\
 &  & \qquad-\frac{-\left(\mu_{-}^{x}\right)^{2}+\left(\mu_{-}^{z}\right)^{2}}{\mu_{-}^{2}}
\sin^{2}\left(\frac{1}{\hbar}\mu_{-}t\right)\Biggr]\,,
\end{eqnarray}
\begin{eqnarray}
\langle\psi_{z}|\sigma^{y}_{H}(t)|\psi_{z}\rangle & = & \frac{\mu_{-}^{z}}{\mu_{-}}\cos\left(\frac{1}{\hbar}\mu_{+}t\right)
\sin\left(\frac{1}{\hbar}\mu_{-}t\right)\nonumber\\
 & + & \frac{\mu_{+}^{z}}{\mu_{+}}\cos\left(\frac{1}{\hbar}\mu_{-}t\right)
\sin\left(\frac{1}{\hbar}\mu_{+}t\right)\,,
\end{eqnarray}
\begin{eqnarray}
\langle\psi_{z}|\sigma^{z}_{H}(t)|\psi_{z}\rangle & = & 
\cos\left(\frac{1}{\hbar}\mu_{+}t\right)\cos\left(\frac{1}{\hbar}\mu_{-}t\right)\nonumber\\
 & + & \frac{\mu_{+}^{x}\mu_{-}^{x}+\mu_{+}^{z}\mu_{-}^{z}}
{\mu_{+}\mu_{-}}\nonumber\\
 & & \cdot\sin\left(\frac{1}{\hbar}\mu_{+}t\right)
\sin\left(\frac{1}{\hbar}\mu_{-}t\right)\,.
\end{eqnarray}

\begin{figure}
\centerline{\includegraphics[width=8cm]{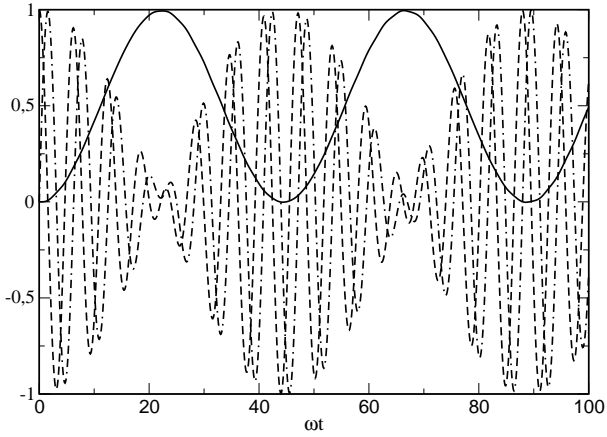}} 
\caption{The spin expectation values $\langle\psi_{z}|\vec \sigma_{H}(t)|\psi_{z}\rangle$ as a function
of time at the resonance condition $2\alpha k_{0y}=\hbar\omega$ with $k_{0y}\lambda=5$
(cf. mid panel of Fig.~\ref{fig1}). The time evolution of
$\langle\psi_{z}|\sigma^{x}_{H}(t)|\psi_{z}\rangle$ (solid line) is governed by a single frequency as shown in 
Eq.~(\ref{approxsigmax}), while the other components
$\langle\psi_{z}|\sigma^{y}_{H}(t)|\psi_{z}\rangle$ and $\langle\psi_{z}|\sigma^{z}_{H}(t)|\psi_{z}\rangle$ (dashed lines)
show beatings between two frequencies given by $\mu_{\pm}$.
\label{fig3}}
\end{figure}

Interestingly the expectation value
$\langle\psi_{z}|\sigma^{x}_{H}(t)|\psi_{z}\rangle$ shows a particular behavior at the resonance $\hbar\omega=\pm2\alpha k_{0y}$.
Here we have $\mu_{\mp}^{z}=0$, and if $\mu_{\pm}^{x}$ can again be neglected compared to $\mu_{\pm}^{z}$
(as it is the case for the choice of parameters used in 
Figs.~\ref{fig1},\ref{fig2}), this expectation value is approximately given by
\begin{equation}
\langle\psi_{z}|\sigma^{x}_{H}(t)|\psi_{z}\rangle\approx\pm\left(1-\cos\left(\frac{2}{\hbar}\mu^{x}_{\mp}t\right)\right)\,.
\label{approxsigmax}
\end{equation}
Thus, the time dependence is, to a very good degree of approximation,
governed by a single frequency. This remarkable
result is also confirmed by numerical 
simulations of the full multiband model shown in Fig.~\ref{fig3}.

Finally, if the electron spin points initially in the $x$-direction, we have
\begin{eqnarray}
\langle\psi_{x}|\sigma^{x}_{H}(t)|\psi_{x}\rangle & = & 
\cos^{2}\left(\frac{1}{\hbar}\mu_{+}t\right)\nonumber\\
 & +  & \frac{-\left(\mu_{+}^{x}\right)^{2}+\left(\mu_{+}^{z}\right)^{2}}{\mu_{+}^{2}}
\sin^{2}\left(\frac{1}{\hbar}\mu_{+}t\right)
\end{eqnarray}
while the expectation values of the other two spin components strictly
vanish. The latter result is also true for the full multi-band model and
follows from the same symmetry considerations
as above.

Let us finally summarize other recent theoretical developments 
pertaining to the issue of {\em zitterbewegung} in quantum wires.
The electron dynamics
in ballistic quantum wires in the presence of spin-orbit coupling were
also recently analyzed by Nikolic, Zarba, and Welack concentrating on 
transverse forces on the electron induced by spin-orbit coupling
\cite{Nikolic05}. A study similar in spirit was carried out by Shen, who
also made a connection between transverse forces due to spin-orbit
interaction and the occurrence of {\em zitterbewegung} \cite{Shen05}.
Lee and Bruder studied recently quantum wires with spin-orbit coupling of
both the Rashba and Dresselhaus type \cite{Lee05}. Their approach is mainly
numerical and concentrates on mapping out charge- and spin-density
modulations in ferromagnet-semiconductor single-junction
quantum wires. The shape of these modulations is explained in terms of
the symmetry properties of the eigenstates of the wire. Finally, Zawadzki
has most recently analysed the band structure of narrow-gap single-wall
carbon nanotubes making a connections to relativistic effects in the free
Dirac equation \cite{Zawadzki05b}. Similarly to
Ref.~\cite{Zawadzki05a}, the kind of {\em zitterbewegung} predicted from
these investigations is along the group velocity of the electron wave
packet, i.e. along the wire, which is considered there as a strictly
one-dimensional system. This is in contrast to the present study where we
consider a quantum wire of finite width and obtain a {\em zitterbewegung}
of electronic wave packets perpendicular to the wire direction.

\section{Holes in a two-dimensional quantum well}
\label{2Dholes}

We now turn to the case of holes in the p-type valence band of III-V 
semiconductors as opposed to s-type
conduction band electrons studied so far. We note
that Jiang {\em et al.} have very recently performed a semiclassical study of
the time evolution of holes in
three-dimensional bulk systems under the influence of a
homogeneous electric field \cite{Jiang05}. This investigation was motivated by 
the recent prediction of the intrinsic spin-Hall effect
\cite{Murakami03,Sinova04,Schliemann05b}. Here we shall analyse the full
quantum time evolution of heavy-hole states in quantum wells, a scenario for
which spin-Hall transport was most recently predicted \cite{Schliemann05b}
and experimentally reported \cite{Wunderlich05}.

At low temperatures and for not too
wide wells, only heavy-hole states are occupied with their angular momentum
pointing predominantly along the growth direction \cite{Winkler00},
corresponding to the total angular momentum states $\pm 3/2$. Due to this
constraint, the effects of structure-inversion asymmetry on the hole spins
are trilinear in the momentum, and the Hamiltonian incorporating this
type of spin-orbit coupling reads for appropriate growth directions
of the quantum well 
\cite{Winkler00,Gerchikov92},
\begin{equation}
{\cal H}=\frac{\vec p^{2}}{2m}+i\frac{\tilde\alpha}{2\hbar^{3}}
\left(p_{-}^{3}\sigma_{+}-p_{+}^{3}\sigma_{-}\right)\,,
\label{defham}
\end{equation}
using the notations $p_{\pm}=p_{x}\pm ip_{y}$, 
$\sigma_{\pm}=\sigma_{x}\pm i\sigma_{y}$, where $\vec p$, $\vec\sigma$
denote the hole momentum operator and Pauli matrices, respectively. 
These Pauli matrices operate on the total angular momentum states
with spin projection $\pm 3/2$ along the growth direction; in this sense they
represent a pseudospin degree of freedom rather than a genuine
spin 1/2. In the above equation, $m$
is the heavy-hole mass, and $\tilde\alpha$ is Rashba spin-orbit coupling 
coefficient due to structure inversion asymmetry. The components of
the time-dependent position operator read
\begin{eqnarray}
x_{H}(t) & = & x(0)+ \frac{p_{x}}{m}t
+\frac{p_{y}}{p^{2}}\frac{3\hbar}{2}\sigma^{z}
\left(1-\cos\left(\frac{2\tilde\alpha p^{3}}{\hbar^{4}}t\right)\right)
\nonumber\\
 & + & \frac{p_{x}}{p^{5}}\frac{3\hbar}{2}
\left(\left(3p_{x}p_{y}^{2}-p_{x}^{3}\right)\sigma^{x}-
\left(3p_{x}^{2}p_{y}-p_{y}^{3}\right)\sigma^{y}\right)\nonumber\\
 & & \cdot\left(\sin\left(\frac{2\tilde\alpha p^{3}}{\hbar^{4}}t\right)
-\frac{2\tilde\alpha p^{3}}{\hbar^{4}}t\right)\nonumber\\
 & + & \frac{\tilde\alpha}{\hbar^{3}}t
\left(6p_{x}p_{y}\sigma^{x}
+3\left(p_{y}^{2}-p_{x}^{2}\right)\sigma^{y}\right)\,,\\
y_{H}(t) & = & y(0)+ \frac{p_{y}}{m}t
-\frac{p_{x}}{p^{2}}\frac{3\hbar}{2}\sigma^{z}
\left(1-\cos\left(\frac{2\tilde\alpha p^{3}}{\hbar^{4}}t\right)\right)\,,
\nonumber\\
 & - & \frac{p_{y}}{p^{5}}\frac{3\hbar}{2}
\left(\left(3p_{x}p_{y}^{2}-p_{x}^{3}\right)\sigma^{x}-
\left(3p_{x}^{2}p_{y}-p_{y}^{3}\right)\sigma^{y}\right)\nonumber\\
 & & \cdot\left(\sin\left(\frac{2\tilde\alpha p^{3}}{\hbar^{4}}t\right)
-\frac{2\tilde\alpha p^{3}}{\hbar^{4}}t\right)\nonumber\\
 & + & \frac{\tilde\alpha}{\hbar^{3}}t
\left(3\left(p_{x}^{2}-p_{y}^{2}\right)\sigma^{x}
+6p_{x}p_{y}\sigma^{y}\right)\,.
\end{eqnarray}
Again, the {\em zitterbewegung} of a wave packet with its spin pointing 
initially 
in the $z$-direction is perpendicular to the group velocity. Specifically,
for an initial state of the form (\ref{initz}) moving along the $y$-direction
($k_{0x}=0$), one finds in the limit
$dk_{0y}\gg 1$ 
\begin{equation}
\langle\psi|x_{H}(t)|\psi\rangle=
\frac{3}{2k_{0y}}\left(1
-\cos\left(\frac{2\tilde\alpha k_{0y}^{3}}{\hbar}t\right)\right)\,.
\end{equation}
Thus, the amplitude of the {\em zitterbewegung} is again proportional to the
wave length $\lambda_{0y}=2\pi/k_{0y}$ of the particle motion perpendicular to it. The 
frequency of the {\em zitterbewegung} is given by 
$\tilde\Omega=2\tilde\alpha k_{0y}^{3}/\hbar$. Winkler {\em et al.} \cite{Winkler02} have
studied both theoretically and experimentally 
the magnitude of the Rashba spin orbit coupling in GaAs-based
quantum well samples with heavy-hole densities of a few $10^{14}{\rm m^{-2}}$
and have found typical values for the characteristic length scale
$m\tilde\alpha/\hbar^{2}$ of a few nanometers. Assuming a value of $2{\rm nm}$ this
corresponds to a coupling parameter of $\tilde\alpha=0.3{\rm eVnm^{3}}$, where we
have used the heavy-hole mass $m\approx 0.5m_{0}$ for GaAs \cite{Vurgaftman01}.
For a typical wave vector with $k_{0y}\approx 0.1{\rm nm}^{-1}$ this leads to 
frequencies $\tilde\Omega$ of order $10^{11}{\rm Hz}$, an estimate
which is of a similar order
of magnitude as for the case of an n-doped quantum well.

\section{Conclusions}
\label{conclusions}

We have investigated the notion of {\em zitterbewegung} in both n- and p-doped
III-V zinc-blende semiconductor quantum wells, extending previous
work on the two-dimensional electron gas \cite{Schliemann05a}. 
{\em zitterbewegung} in the two-dimensional electron gas has been 
studied for the
case of spin-orbit coupling of both the Rashba and Dresselhaus type. In the
context of these investigations we have also discussed recent work by
Zawadzki \cite{Zawadzki05a}. The crucial difference between 
{\em zitterbewegung} of free electrons and electrons bound in the above
semiconductor nanostructures is the fact that in the latter systems the
frequency of the oscillations is by orders of magnitude smaller while the
amplitude is grossly larger. This circumstances make such systems favorable
candidates for the experimental detection of {\em zitterbewegung}.

The case of electron dynamics in a quantum wire is studied in great detail for
various initial conditions. For this type of system we propose possible
experiments for detecting the {\em zitterbewegung} of electronic wave packets.
For an harmonic quantum well in the presence of Rashba spin-orbit coupling,
the dynamical parameters can be tuned to a resonance condition where the
amplitude of the {\em zitterbewegung} becomes maximal \cite{Schliemann05a}. 
This property should
facilitate the experimental observation of this effect. In addition to the
orbital dynamics, we have also analyzed in detail
the electron spin dynamics, which
also show peculiarities at the resonance.

Finally we have also discussed in detail the {\em zitterbewegung} in the
two-dimensional hole gas. Here we have considered spin-orbit coupling
of heavy holes due to structure-inversion asymmetry. Concerning the
frequency and amplitude of the {\em zitterbewegung}, similar results are
obtained as for the two-dimensional electron gas.

\acknowledgments{We thank E.~S. Bernardes, T. Dekorsy, and J.~C. Egues for
useful discussions. The work of D.~L. and R.~M.~W. was supported by DARPA.
D.~L also acknowledges support from 
the Swiss NSF, the NCCR Nanoscience, EU RTN Spintronics, and ONR.}



\begin{references}
\bibitem{noteerratum}
Eq.~(\protect{\ref{xHexpansion}}) corrects an error in Eq.~(8) of  
Ref.~\cite{Schliemann05a}, which has no influence on further results
presented there.
\end{references}
\end{document}